\documentclass[useAMS,usenatbib]{mn2e} 

\usepackage{graphicx,fleqn}
\usepackage{rotating}
\DeclareGraphicsExtensions{.eps,.ps,.eps.gz,.ps.gz,.eps.Z}
\title[Dust and Gas in the Local Environments of Gamma-Ray Bursts]{Dust and Gas in the Local Environments of Gamma-Ray Bursts}

\author[P.~Schady, K.O.~Mason, M.J.~Page, et al.]{P.~Schady$^{1,2}$, K.O.~Mason$^{3,1}$, M.J.~Page$^{1}$, M.De~Pasquale$^{1}$, D.C.~Morris$^{2}$, P.~Romano$^{4}$, \and P.W.A.~Roming$^{2}$, S.~Immler$^{5}$ and D.E.~Vanden~Berk$^{2}$\\
$^{1}$ The UCL Mullard Space Science Laboratory, Holmbury St Mary, Dorking, Surrey, RH5 6NT, UK.\\
$^{2}$ Department of Astronomy and Astrophysics, Pennsylvania State University, 525 Davey Laboratory, University Park, PA 16802, USA.\\
$^{3}$ The Particle Physics and Astronomy Research Council, Polaris House, North Star Avenue, Swindon, Wiltshire, SN2 1SZ, UK.\\
$^{4}$ INAF-Osservatorio Astronomico di Brera, via E. Bianchi 46, 23807 Merate (LC), Italy.\\
$^{5}$ NASA/Goddard Space Flight Center, Greenbelt, MD 20771, USA.
}

\date{Received: }

\begin{document}

\newcommand\eg{e.g. } 
\newcommand\ie{i.e. } 
\newcommand\swift{{\it Swift}} 
\newcommand\sqrcm{cm$^{2}$} 
\newcommand\invsqrcm{cm$^{-2}$} 
\newcommand\flu{$\mathrm{erg}~\mathrm{cm}^{-2}$}
\newcommand\flux{$\mathrm{erg}~\mathrm{s}^{-1}$}
\newcommand\cps{$\mathrm{counts}~\mathrm{s}^{-1}$}
\newcommand\nh{$N_{H}$}
\newcommand\nhx{$N_{H,X}$}
\newcommand\av{$A_V$}
\newcommand\Eiso{$E_{iso}$}
\def\lesssim{\mathrel{\hbox{\rlap{\hbox{\lower4pt\hbox{$\sim$}}}\hbox{$<$}}}}
\def\gtrsim{\mathrel{\hbox{\rlap{\hbox{\lower4pt\hbox{$\sim$}}}\hbox{$>$}}}}

\maketitle 

\begin{abstract}
 Using a sample of gamma-ray burst (GRB) afterglows detected by both the X-Ray and the UV/Optical Telescopes (XRT and UVOT) on \swift, we modelled the spectral energy distributions (SEDs) to determine gas column densities and dust extinction in the GRB local environment. In six out of seven cases we find an X-ray absorber associated with the GRB host galaxy with column density (assuming solar abundances) ranging from $(0.8 - 7.7)\times 10^{21}$~\invsqrcm. We determine the rest-frame visual extinction \av\ using the SMC, LMC and Galactic extinction curves to model the dust in the GRB host galaxy, and this ranges from \av\ = $0.12\pm 0.04$ to \av\ = $0.65^{+0.08}_{-0.07}$. The afterglow SEDs were typically best fit by a model with an SMC extinction curve. In only one case was the GRB afterglow better modelled by a Galactic extinction curve, which has a prominent absorption feature at 2175~\AA. We investigate the selection effects present in our sample and how these might distort the true distribution of \av\ in GRB host galaxies. We estimate that GRBs with no afterglow detected blueward of 5500~\AA\ have average rest-frame visual extinctions almost eight times those observed in the optically bright population of GRBs. This may help account for the $\sim 1/3$ of GRBs observed by \swift\ that have no afterglow detected by UVOT.
\end{abstract} 


\section{Introduction} 
There is now a wealth of observational evidence linking long duration gamma-ray bursts (GRBs) (prompt emission lasting $> 2$~s) with the collapse of a massive star~\citep[collapsar model;~e.g.][]{woo93}. This includes the underlying supernova features in the afterglow of some GRBs~[\eg~GRB~980425 \citep{kfw+98}, GRB~030329 \citep{hsm+03} and GRB~060218 \citep{cmb+06}], and the association between GRB host galaxies and high-mass star formation \citep{tbb+04}. The $\gamma$-ray energy emission of GRBs is unaffected by dust and gas in the intervening interstellar medium, which combined with the vast energy released during the initial explosion, allows them to be detected out to very high redshifts \citep[e.g.][]{tac+05}. These two facts give GRBs the potential to be highly powerful tools with which to trace the star formation history (SFH) in an unbiased way. Furthermore, the longer lived, lower energy afterglows light up their host galaxies, albeit for only a brief time (on the order of weeks), providing invaluable insight into the chemical makeup of these galaxies that would otherwise be unattainable in a majority of cases; certainly at high redshifts.

To fully maximise the potential that GRBs offer as cosmic probes the selection effects present in GRB studies need to be known, such as the stellar populations that they trace and the properties of their host galaxies. Infra-red observations of long GRB host galaxies taken with {\it Spitzer} indicate that they are not strong starburst galaxies, nor are they particularly dusty \citep{lcf+06}. This is in conflict with expectations from the collapsar model, which predicts GRBs to occur in regions of active star formation that are heavily enshrouded by dust. This mismatch between observations and theory could very possibly be the result of selection effects, whereby the near infra-red (NIR) to ultra violet (UV) afterglow of GRBs in dusty galaxies are extinguished, thus reducing the chance of identifying the host. Further detailed analysis of GRB environments is important to determine the range in host galaxy properties, and, fundamentally, to provide a better understanding of our observations and the limitations that they present.

Due to the broadband power law spectral behaviour of GRB afterglows, the effects of absorbing dust and gas in the local environment on the GRB spectral energy distribution (SED) can be well identified. In the analysis of X-ray and optical afterglow spectra for eight GRBs, \citet{gw01} found evidence for high column densities of gas in the GRB local environment that were comparable with those observed in giant molecular clouds. Furthermore, they found the optical extinction to be 10-100 times smaller than expected given the column densities. \citet{sfa+04} extended this sample and found the ratio of host galaxy column density to visual extinction to be an order of magnitude larger that that observed in the Milky Way (MW), and also greater than that in the Small Magellanic Clouds (SMC) and Large Magellanic Clouds (LMC). Prompted by theoretical studies that indicate that the intense radiation emitted by a GRB should destroy small dust grains out to radii of around $10$~pc [\eg~Fruchter, Krolik \& Rhoads (2001); Perna \& Lazzati (2002); Perna, Lazzati \& Fiore (2003)], the large gas-to-dust ratio was taken to be evidence of the destruction of dust in the surrounding vicinity of the burst, thus reducing the visual extinction observed.

\citet{sfa+04} also found little evidence of the strong 2175~\AA\ Galactic absorption feature in their optical-NIR spectral analysis. Instead they found the SED to be best fit by a model in which the host galaxy has an SMC or starburst galaxy dust extinction law, which has no such absorption feature. More recently Kann, Klose \& Zeh (2006) analysed the afterglow spectral energy distributions in the optical and NIR bands on a larger sample of 30 pre-\swift\ GRBs, and also found the SMC extinction curve to provide a better fit to the spectra than the MW or LMC extinction curves.

The simultaneous observations taken with the X-Ray Telescope [XRT; Burrows et al. (2005a)] and UV/Optical Telescope [UVOT; Roming et al. (2005)] on-board the \swift\ spacecraft \citep{gcg+04} provide GRB afterglow SEDs without the need to extrapolate data to the same epoch. This capability is unique to \swift\ and the accurate broadband spectral modelling that is possible with this provides an important data set to compare to previous multi-wavelength GRB samples. In this paper we model the SEDs of a sample of \swift\ GRBs with afterglows detected by both the XRT and UVOT to investigate the rest-frame visual extinction and soft X-ray absorption at the GRB host galaxy. This provides an indication of the dust and gas content in the local environment of the GRB, which we compare to the environment of the MW, the LMC and the SMC.

The effects of the selection bias that is introduced by analysing only those GRBs with X-ray and optical afterglows is addressed by using our results to examine the dust in the local environments of GRBs that have no UVOT detected afterglow. Our results have implications for the role that dust plays in accounting for the lack of an UV/optical afterglow detection in $\sim 1/3$ of GRBs observed by the UVOT.

In section~\ref{sec:anlys} we present the GRB sample and describe the X-ray and UV/optical data reduction and analysis and in section~\ref{sec:mod} we describe the models that we used to fit the data. We present the results from our spectral modelling and discuss their implications in sections~\ref{sec:rslts} and \ref{sec:disc}, and also investigate how GRBs that have no UV/optical afterglow detected by UVOT may be affected by dust. Our conclusions are summarised in section~\ref{sec:conc}. Throughout the paper temporal and spectral indices, $\alpha$ and $\beta$ respectively, are denoted such that $F_t\propto t^{-\alpha}$ and $F_{\nu}\propto\nu^{-\beta}$, and all errors are 1$\sigma$ unless specified otherwise.

\section{Data Reduction and Analysis}\label{sec:anlys}
SEDs at a single-epoch were produced for a total of seven bursts. This epoch was chosen to be 1~hour after the onset of the initial prompt emission (time T) for four of the bursts (GRB~050318, GRB~050525, GRB~050802 and GRB~060512). For the remaining three GRBs (GRB~050824,GRB~051111 and GRB~060418) the SEDs were produced at a time T+2hrs due to a lack of data at T+1hr, resulting from spacecraft observing constraints.

The GRBs in our sample were chosen on the basis that they had afterglows detected by both the XRT and UVOT instruments and a spectroscopic redshift with $z < 1.75$. Bursts for which the photometry was considered to be too poor to provide useful constraints on the spectral fitting were not used. This included those with UVOT detections in fewer that three filters, or bursts that did not have well enough sampled light curves with which to obtain reliable multi-band photometry at a single epoch. To avoid any contamination from absorption caused by the Ly$\alpha$ forest only UVOT data with a rest-frame wavelength $\lambda > 1215$~\AA\ were used in the spectral analysis. The redshift limit was, therefore, a necessary requirement to ensure that there were sufficient optical and UV data points in the afterglow SED to constrain the spectral fitting. At the redshift of our sample the afterglow spectra cover the wavelength range where the redshifted 2175~\AA\ extinction bump is expected to lie, and those GRBs closer to the redshift upper limit also probe the rest-frame far-ultraviolet (FUV) spectra, where the divergence between extinction curves is greatest.

\subsection{UVOT Data}\label{subsec:opt}
\begin{table}
\begin{center}
\caption{UVOT Data\label{tab:UVOTinfo}}
\begin{tabular}{@{}llll}
\hline
GRB & $z$ & Galactic E(B-V)$^{\ 1}$ & $\alpha_{\rm{opt}}$\\
\hline\hline
050318 & 1.44$^{\ 2}$ & 0.017 & 0.94$^{\ 9}$ \\
050525 & 0.606$^{\ 3}$ & 0.095 & 1.56,0.62$^{\ 10}$ \\
050802 & 1.71$^{\ 4}$ & 0.021 & 0.82$^{\ 11}$ \\
050824 & 0.83$^{\ 5}$ & 0.034 & 0.55$^{\ 12}$ \\
051111 & 1.549$^{\ 6}$ & 0.162 & 0.96$^{\ 13}$ \\
060418 & 1.49$^{\ 7}$ & 0.224 & 1.25 \\
060512 & 0.443$^{\ 8}$ & 0.014 & 0.84 \\
\hline\\
\end{tabular}
\end{center}
$^{1}${\footnotesize \ \cite{sfd98}}\\
$^{2}${\footnotesize \ \citet{bm05}};
$^{3}${\footnotesize \ \citet{fcb+05}};
$^{4}${\footnotesize \ \citet{fsj+05}};
$^{5}${\footnotesize \ \citet{fjs+05}};
$^{6}${\footnotesize \ \citet{pro05}};
$^{7}${\footnotesize \ \citet{dfp+06}};
$^{8}${\footnotesize \ \citet{bfk+06}}\\
$^{9}${\footnotesize \ \citet{srm+05}}\\
$^{10}${\footnotesize \ Light curve best fit by double power law~\citep{bbb+06}}
$^{11}${\footnotesize \ \citet{odp+06}}\\
$^{12}${\footnotesize \ \citet{hm05}}\\
$^{13}${\footnotesize \ \citet{blp+06}}\\
\end{table}

\begin{table}
\begin{center}
\caption{X-Ray Data}\label{tab:Xraydata}
\begin{tabular}{@{}lllll}
\hline
GRB & Data & Gal. \nh\ & Start & Exp. \\
 & Mode & ($10^{20}$~\invsqrcm) & Time (s)$^{\ 1}$ & Time (s)\\
\hline\hline
050318 & PC & 2.8 & 3280 & 23431 \\
050525 & PC & 9.1 & 7056 & 3560 \\
050802 & PC & 1.8 & 480 & 4835 \\
050824 & PC & 3.6 & 6096 & 17157 \\
051111 & PC & 5.0 & 5552 & 1583 \\
060418 & WT & 9.2 & 448 & 1841 \\
060512 & PC & 1.4 & 3680 & 58966 \\
\hline
\end{tabular}
\end{center}
$^{1}${\footnotesize \ Measured from the BAT trigger time.}
\end{table}

The UVOT contains three optical and three ultra-violet lenticular filters, which cover the wavelength range between 1600~\AA\ and 6000~\AA. Photometric measurements were extracted from the UVOT imaging data using a circular source extraction region with a $6\arcsec$ radius for the {\it V}, {\it B} and {\it U} optical filters and a $12\arcsec$ radius for the three UVOT ultra-violet filters to remain compatible with the current effective area calibrations\footnotemark[1]. 
\footnotetext[1]{http://heasarc.gsfc.nasa.gov/docs/heasarc/caldb/swift/docs/uvot/}
Where possible the background rate was taken from an annulus with $12\arcsec$ inner radius and $20\arcsec$ outer radius centred on the source. In the cases where there were nearby sources that contaminated this extraction region, the background was taken from a source-free region close to the target with radii ranging between $10$ and $20\arcsec$.

For each of the UVOT lenticular filters the tool {\sc uvot2pha} (version 1.1) was used to produce spectral files compatible with {\sc xspec} (version 12.2.1), and response matrices were taken from version 102 of the UVOT calibration files. To create an SED at an instantaneous epoch the count rates in these files were set to correspond to the count rate of the GRB at the appropriate epoch. These count rates and the associated errors were determined from power law model fits to the light curve in each filter where the GRB afterglow is assumed to decay at the same rate in the UV and optical bands. This is justified by the small variations present in the decay rate between filters for the GRBs in our sample, which have a typical variance of $\sim 0.2$ and are consistent within $1\sigma$ errors. Furthermore, the colour evolution expected when the cooling break migrates through the optical bands is typically observed at later times than the epochs we deal with, on the order of $\sim 10^4$~s \citep[e.g.][]{bbb+06}.

Where possible the UV/optical decay index was taken from the literature where UVOT data were used in the analysis. Otherwise, this analysis was carried out by ourselves, with the exception of GRB~050824 and GRB~051111, for which there was not sufficient UVOT data to constrain the fits to the light curves and, therefore, reported indices from $R$-band light curve fits were used. For the analysis done on UVOT data, both our own and that reported in the literature, the decay index was determined from the combined UVOT light curve, which was produced by normalising each filter to the $V$-band light curve. The count rate at an instantaneous epoch for each filter was then determined from the best-fit temporal decay model using the appropriate normalisation. The light curves were modelled as a power law, with the exception of GRB~050525, which required further components to describe the more complex early time temporal behaviour \citep{bbb+06}. The decay rates determined for each burst are listed in Table~\ref{tab:UVOTinfo}, as well as the references used where appropriate.

\begin{figure}
\centering
\includegraphics[width=0.5\textwidth]{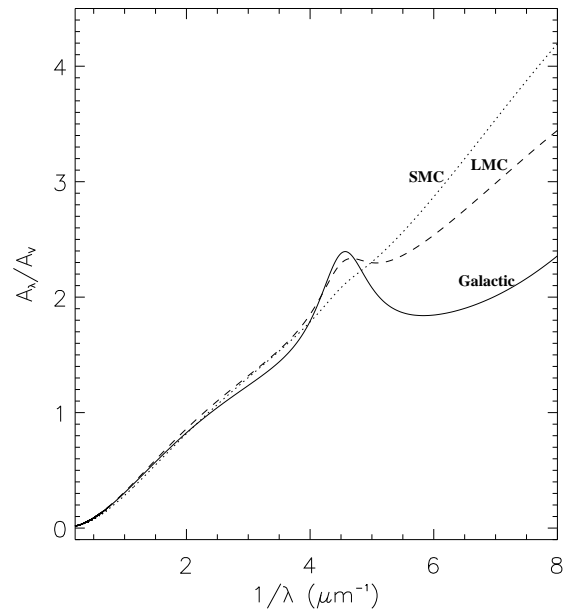}
\caption{Galactic (solid), LMC (dashed) and SMC (dotted) extinction curves. Parameterisations taken from \citet{pei92}}\label{fig:extcurves}
\end{figure}

\subsection{X-Ray Data}\label{subsec:xray}
The XRT covers the 0.2 -- 10~keV energy range, and all data were reduced using the {\sc xrtpipeline} tool (version 0.9.9). In most cases the data used were taken in photon counting (PC) mode with the exception of GRB~060418, for which the majority of the data taken at the time of interest are in window timing (WT) mode \citep{hbn+04}. For PC data source counts were extracted from a circular region centred on the source with an outer radius ranging from $50\arcsec$ to $95\arcsec$. In the case of GRB~050802 the data suffered from pile-up and, therefore, an annular extraction region was used to exclude those pixels that were piled-up. The inner radius to this was $9.44\arcsec$ (4 pixels) and the outer radius was $106.2\arcsec$ (45 pixels). The background count rate was estimated from a circular, source-free area in the field of view (FOV) with a radius of $118\arcsec$ (50 pixels). For WT mode data, the extraction regions used for the source and background count rates were $94\arcsec$ slits positioned over the source and in a source free region of the FOV, respectively. {\sc xselect} (version 2.4) was used to extract spectral files from the event data in the energy ranges 0.3 -- 10~keV for PC and WT mode data, which is the recommended band to use for compatibility with the current calibration files \footnotemark[2]. 
\footnotetext[2]{http://heasarc.gsfc.nasa.gov/docs/heasarc/caldb/swift/docs/xrt/}
Corresponding effective area files were created using the {\sc xrtmkarf} tool (version 0.5.1) and binned to at least 20 counts per energy bin. Response matrices from version 8 of the XRT calibration files were used for both WT and PC mode data.

A large proportion of GRBs observed by XRT have shown spectral evolution in the first few thousand seconds of emission, during X-ray flares and after early time temporal breaks, which could be the transition from internal shock to external shock dominated emission \citep[e.g][]{brf+05,zfd+06}. All X-ray data were taken from time intervals where spectral evolution was no longer observed, and the spectra were normalised to the epoch corresponding to the SED by using the best-fit model to the X-ray light curve, in the same way as for the UVOT data. The data mode and time intervals used are listed in Table~\ref{tab:Xraydata}.

\section{The Model}\label{sec:mod}
To model the afterglow spectral continuum we tried both a power law and broken power law fit, where in the latter the change in spectral slope was fixed to $\Delta\beta =0.5$ to correspond to the change in slope caused by a cooling break. In addition to this a constraint was also imposed on the break energy such that it was within the observing window (\ie\ $0.002~\rm{keV}<E_b<10.0$~keV). Any fit with a spectral break outside this energy range would be equivalent to a power law fit to the data. In both the power law and broken power law models we included two dust and gas components to correspond to the Galactic and the host galaxy photoelectric absorption and dust extinction. The Galactic column density and reddening in the line of sight were fixed to the values taken from \citet{dl90} and Schlegel, Finkbeiner \& Davis (1998), respectively. The second photoelectric absorption system was set to the redshift of the GRB, and the equivalent neutral hydrogen column density in the host galaxy was determined assuming solar abundances. The dependence of dust extinction on wavelength in the GRB host galaxy was modelled on the empirical extinction laws corresponding to the MW, LMC and SMC.

The wavelength dependence on dust extinction observed in these three environments is well reproduced by a dust model composed of silicate and graphite grains, where variations in the relative abundance and grain size distribution produce the different extinction laws \citep{pei92}. The parameterisation of the dust extinction laws in the MW, LMC and SMC are shown in Fig.~\ref{fig:extcurves}, where $R_{V}=A_{V}/E(B-V)=3.08$, 2.93 and 3.16 for the Galactic, SMC and LMC extinction laws, respectively \citep{pei92}.

The prominence of the 2175~\AA\ absorption feature and amount of FUV extinction vary between the three curves. The MW has the strongest extinction at 2175~\AA\ and smallest amount of FUV extinction, whereas the SMC has the greatest amount of FUV extinction, rising faster than $1/\lambda$, and an insignificant 2175~\AA\ feature. A further difference in these three extinction laws is in the amount of reddening per H atom \citep{dra00}, which is observed to be greatest in the MW and least in the SMC.

We chose these three curves to model the rest frame visual extinction in the GRB afterglows to identify the predominant extinction properties of the dust within the GRB local environment. We refer to each of the spectral models as the MW, SMC and LMC model, where the name corresponds to the extinction law used to describe the dust extinction properties in the GRB host galaxy.

\begin{figure}
\centering
\includegraphics[width=0.5\textwidth]{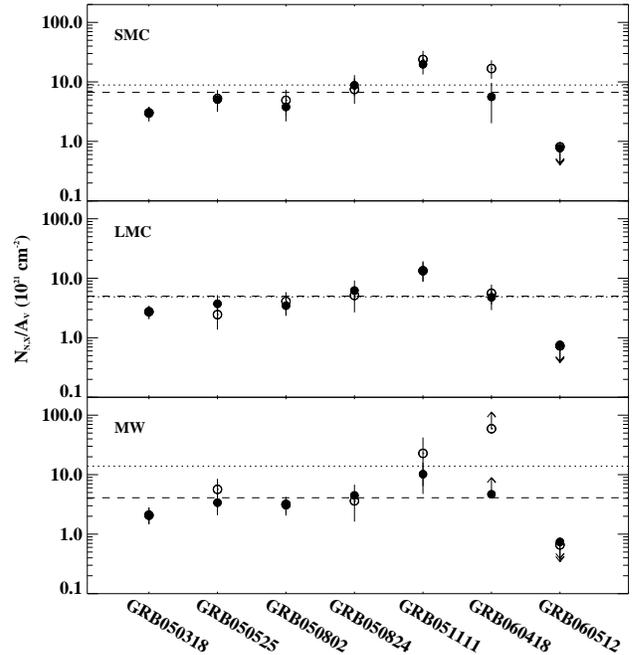}
\caption{Ratio of rest-frame \nhx\ to \av\ for the three models used in the spectral analysis; SMC model (top panel), LMC model (middle panel) and Milky Way model (bottom panel). Filled circles correspond to power law spectral models and open circles to broken power law fits. Dashed and dotted lines show the mean \nhx/\av\ value determined from each spectral model for a power law and broken power law fit, respectively.}\label{fig:rat}
\end{figure}

\begin{table*}
\centering
\begin{minipage}{126mm}
\caption{Results from simultaneous UV/optical and X-ray power law spectral fits, where only UVOT data with rest-frame wavelength $\lambda > 1215$~\AA were used.\label{tab:powsedfits}}
\begin{tabular}{@{}ccccccc}
\hline
GRB & Model & \nhx$^{\ a}$ & $\beta$ & \av$^{\ a}$ & $\chi ^2$ & Null Hypothesis\\
& & $10^{21}$~\invsqrcm & & & (dof) & Probability\\
\hline\hline
050318 & SMC & $1.56^{+0.44}_{-0.42}$ & $0.95\pm 0.03$ & $0.53\pm 0.06$ & 101 (76) & 0.03\\
 & LMC & $2.28^{+0.53}_{-0.51}$ & $1.04\pm 0.04$ & $0.83^{+0.10}_{-0.09}$ & 105 (76) & 0.02\\
 & MW & $1.90^{+0.60}_{-0.55}$ & $0.99^{+0.06}_{-0.05}$ & $0.91\pm 0.14$ & 154 (76) & 0.00\\
\hline
050525 & SMC & $0.78^{+0.32}_{-0.29}$ & $0.76\pm 0.01$ & $0.16\pm 0.03$ & 29 (33) & 0.67\\
 & LMC & $0.87^{+0.32}_{-0.29}$ & $0.79\pm 0.02$ & $0.23\pm 0.04$ & 33 (33) & 0.48\\
 & MW & $0.78^{+0.33}_{-0.30}$ & $0.78\pm 0.03$ & $0.23\pm 0.06$ & 58 (33) & 0.01\\
\hline
050802 & SMC & $0.91^{+0.41}_{-0.39}$ & $0.66\pm 0.02$ & $0.24\pm 0.03$ & 106 (81) & 0.03\\
 & LMC & $1.30^{+0.44}_{-0.42}$ & $0.70\pm 0.02$ & $0.38\pm 0.04$ & 98 (81) & 0.10\\
 & MW & $1.98^{+0.50}_{-0.48}$ & $0.76\pm 0.03$ & $0.65^{+0.08}_{-0.07}$ & 88 (81) & 0.29\\
\hline
050824 & SMC & $1.08^{+0.42}_{-0.37}$ & $0.95\pm 0.02$ & $0.12\pm 0.04$ & 27 (24) & 0.33\\
 & LMC & $1.14^{+0.43}_{-0.38}$ & $0.96\pm 0.03$ & $0.18\pm 0.05$ & 27 (24) & 0.29\\
 & MW & $1.20^{+0.45}_{-0.38}$ & $0.98\pm 0.04$ & $0.27\pm 0.10$ & 31 (24) & 0.16\\
\hline
051111 & SMC & $7.71^{+1.80}_{-1.55}$ & $1.10\pm 0.06$ & $0.39^{+0.11}_{-0.10}$ & 13 (11) & 0.28\\
 & LMC & $8.50^{+1.99}_{-1.73}$ & $1.18^{+0.09}_{-0.08}$ & $0.64^{+0.19}_{-0.18}$ & 14 (11) & 0.23\\
 & MW & $8.63^{+2.29}_{-2.01}$ & $1.19\pm 0.13$ & $0.85^{+0.35}_{-0.34}$ & 22 (11) & 0.03\\
\hline
060418 & SMC & $0.97^{+0.69}_{-0.62}$ & $0.89\pm 0.01$ & $0.17\pm 0.02$ & 79 (75) & 0.36\\
 & LMC & $1.83^{+0.76}_{-0.68}$ & $0.97\pm 0.02$ & $0.38\pm 0.05$ & 68 (75) & 0.70\\
 & MW & $0.72^{+0.76}_{-0.71}$ & $0.88\pm 0.04$ & $0.15\pm 0.10$ & 156 (75) & 0.13\\
\hline
060512 & SMC & $< 0.34$ & $0.99\pm 0.02$ & $0.44^{+0.04}_{-0.05}$ & 33 (20) & 0.04\\
 & LMC & $< 0.33$ & $0.98\pm 0.02$ & $0.44\pm 0.05$ & 37 (20) & 0.01\\
 & MW & $< 0.27$ & $0.96\pm 0.02$ & $0.37\pm 0.04$ & 44 (20) & 0.002\\
\hline
\end{tabular}
\\
$^{a}${\footnotesize \ At the redshift of the GRB}
\end{minipage}
\end{table*}

\begin{table*}
\centering
\begin{minipage}{126mm}
\caption{Simultaneous UV/optical and X-ray broken power law spectral fits. Only using UVOT data with rest-frame wavelength $\lambda > 1215$~\AA.\label{tab:bknpsedfits}}
\begin{tabular}{@{}ccccccccccc}
\hline
GRB & Model & \nhx$^{\ a}$ & $\beta_1$ & $E_b$ & $\beta_2$ & \av$^{\ a}$ & $\chi ^2$ & Null Hypothesis\\
& & $10^{21}$~\invsqrcm & & (keV) & & & (dof) & Probability\\
\hline\hline
050318 & SMC & $1.70^{+0.42}_{-0.34}$ & $0.47^{+0.03}_{-0.02}$ & $0.003\pm 0.0003$ & $0.97^{+0.04}_{-0.03}$ & $0.56^{+0.03}_{-0.04}$ & 99 (75) & 0.03\\
 & LMC & $2.28^{+0.41}_{-0.44}$ & $0.54\pm 0.04$ & $0.002 ^{+0.001}_{-0.0}$ & $1.04\pm 0.04$ & $0.83^{+0.09}_{-0.08}$ & 105 (75) & 0.01\\
 & MW & $1.89^{+0.51}_{-0.48}$ & $0.50\pm 0.05$ & $< 0.0004$ & $1.00\pm 0.05$ & $0.91^{+0.13}_{-0.14}$ & 154 (75) & $2e^{-7}$\\
\hline
050525 & SMC & $1.39^{+0.48}_{-0.51}$ & $0.42^{+0.11}_{-0.10}$ & $0.029 ^{+0.104}_{-0.022}$ & $0.92^{+0.11}_{-0.10}$ & $0.26\pm 0.04$ & 27 (32) & 0.73\\
 & LMC & $0.92^{+0.30}_{-0.37}$ & $0.30^{+0.01}_{-0.03}$ & $0.004 ^{+0.005}_{-0.001}$ & $0.80^{+0.01}_{-0.03}$ & $0.38^{+0.03}_{-0.07}$ & 32 (32) & 0.47\\
 & MW & $1.36^{+0.57}_{-0.40}$ & $0.67^{+0.03}_{-0.04}$ & $1.048 \pm 0.290$ & $1.17^{+0.03}_{-0.04}$ & $0.24\pm 0.07$ & 80 (32) & $5e^{-6}$\\
\hline
050802 & SMC & $0.89^{+0.42}_{-0.40}$ & $0.61\pm 0.02$ & $2.970 ^{+0.286}_{-0.348}$ & $1.11\pm 0.02$ & $0.18\pm 0.03$ & 84 (80) & 0.36\\
 & LMC & $1.14^{+0.44}_{-0.43}$ & $0.64\pm 0.03$ & $2.990 ^{+0.516}_{-0.246}$ & $1.14\pm 0.03$ & $0.28^{+0.05}_{-0.05}$ & 81 (80) & 0.43\\
 & MW & $1.73^{+0.51}_{-0.49}$ & $0.72^{+0.03}_{-0.05}$ & $3.960 ^{+0.538}_{-0.900}$ & $1.22^{+0.03}_{-0.05}$ & $0.55^{+0.08}_{-0.11}$ & 79 (80) & 0.53\\
\hline
050824 & SMC & $1.16^{+0.46}_{-0.37}$ & $0.47^{+0.05}_{-0.03}$ & $0.003 \pm 0.001$ & $0.97^{+0.05}_{-0.03}$ & $0.16^{+0.06}_{-0.04}$ & 26 (23) & 0.32\\
 & LMC & $1.27^{+0.45}_{-0.44}$ & $0.50^{+0.03}_{-0.04}$ & $0.003 \pm 0.001$ & $1.00^{+0.03}_{-0.04}$ & $0.25^{+0.05}_{-0.08}$ & 27 (23) & 0.28\\
 & MW & $1.36^{+0.48}_{-0.36}$ & $0.53^{+0.03}_{-0.04}$ & $0.003 \pm 0.001$ & $1.03^{+0.03}_{-0.04}$ & $0.38^{+0.03}_{-0.18}$ & 30 (23) & 0.14\\
\hline
051111 & SMC & $10.04^{+2.70}_{-2.79}$ & $0.82^{+0.24}_{-0.20}$ & $0.040 ^{+0.820}_{-0.034}$ & $1.32^{+0.24}_{-0.20}$ & $0.42\pm 0.13$ & 12 (10) & 0.25\\
 & LMC & $8.50^{+1.72}_{-1.63}$ & $0.68^{+0.42}_{-0.07}$ & $< 0.772$ & $1.180^{+0.42}_{-0.07}$ & $0.64^{+0.26}_{-0.18}$ & 14 (10) & 0.17\\
 & MW & $11.45^{+2.56}_{-2.83}$ & $0.95^{+0.18}_{-0.10}$ & $< 0.883$ & $1.45^{+0.18}_{-0.10}$ & $0.50^{+0.41}_{-0.34}$ & 23 (10) & 0.01\\
\hline
060418 & SMC & $2.80^{+0.97}_{-0.85}$ & $0.85\pm 0.01$ & $1.279 ^{+0.254}_{-0.187}$ & $1.35\pm 0.01$ & $0.17\pm 0.02$ & 68 (74) & 0.68\\
 & LMC & $2.05^{+0.78}_{-0.71}$ & $0.95\pm 0.02$ & $2.100 ^{+0.684}_{-0.513}$ & $1.45\pm 0.02$ & $0.37\pm 0.05$ & 64 (74) & 0.79\\
 & MW & $1.64^{+0.93}_{-0.89}$ & $0.79^{+0.02}_{-0.01}$ & $1.347 ^{+0.526}_{-0.150}$ & $1.29^{+0.02}_{-0.01}$ & $< 0.03$ & 145 (74) & $2e^{-6}$\\
\hline
060512 & SMC & $< 0.47$ & $0.54^{+0.05}_{-0.01}$ & $0.003 \pm 0.001$ & $1.04^{+0.05}_{-0.01}$ & $0.58^{+0.06}_{-0.03}$ & 32 (19) & 0.04\\
 & LMC & $< 0.42$ & $0.53\pm 0.01$ & $0.003 \pm 0.001$ & $1.03\pm 0.01$ & $0.57^{+0.06}_{-0.03}$ & 35 (19) & 0.01\\
 & MW & $< 0.32$ & $0.51\pm 0.01$ & $< 0.001$ & $1.010\pm 0.01$ & $0.49^{+0.04}_{-0.003}$ & 44 (19) & 0.001\\
\hline
\end{tabular}
$^{a}${\footnotesize \ At the redshift of the GRB}\\
\end{minipage}
\end{table*}

\begin{figure*}
\centering
\includegraphics[width=1.0\textwidth]{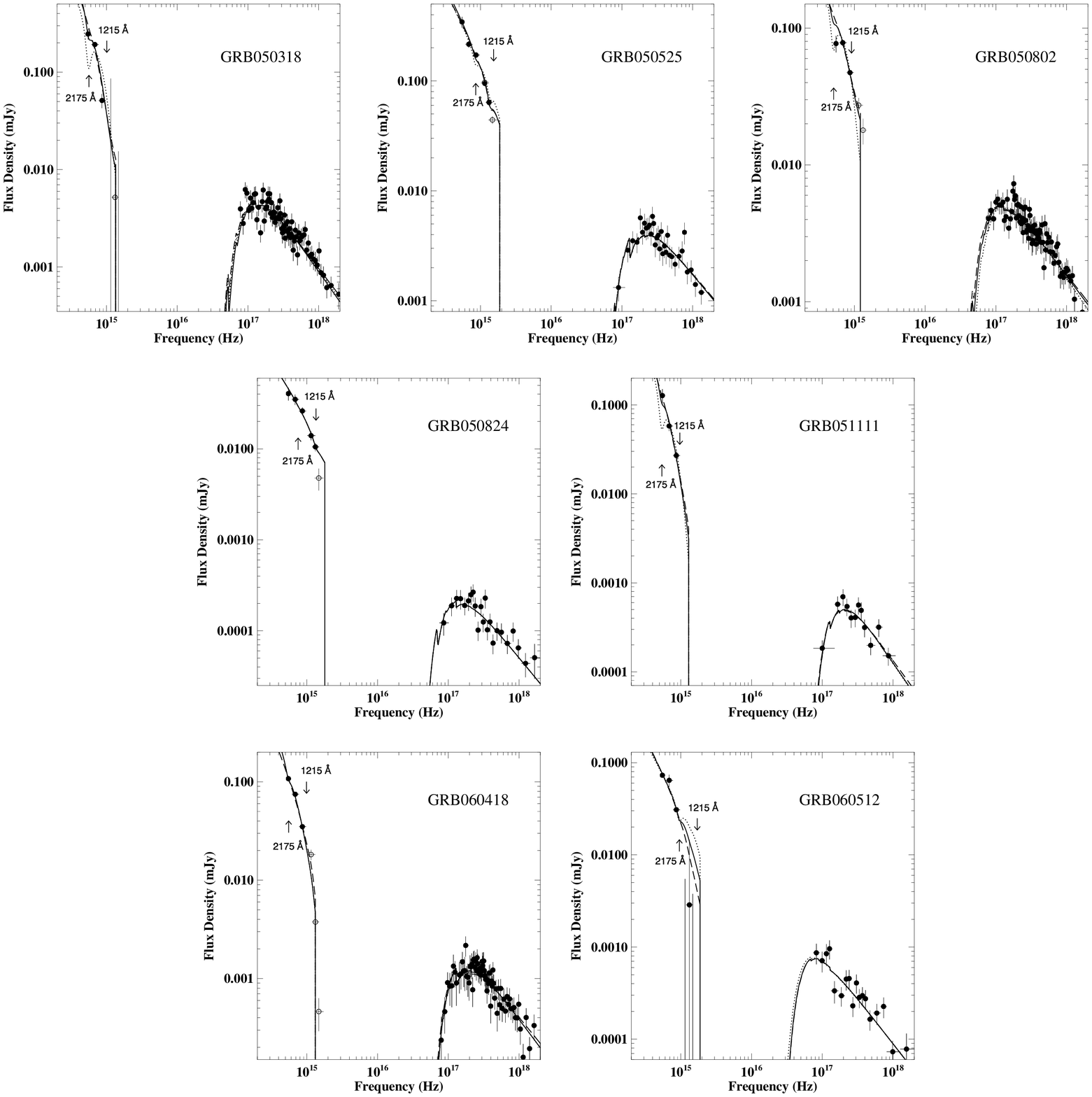}
\caption{SEDs for seven GRBs at an instantaneous epoch (see text) with best-fit models for each corresponding dust extinction curve shown; SMC (dashed), LMC (solid) and Galactic (dotted). Open circles are data points at wavelength $\lambda < 1215$~\AA\ in the rest frame, and therefore not used in the spectral fitting. The positions of 2175~\AA\ and 1215~\AA\ in the rest frame are indicated.}\label{fig:seds}
\end{figure*}

\section{Results}\label{sec:rslts}
The value that we determine in our spectral analysis for \nh\ is an equivalent neutral hydrogen column density that results from the amount of soft X-ray absorption in the spectrum, where solar abundances are assumed. This is dominated primarily by oxygen K-shell absorption \citep{mm83}. To distinguish between the equivalent neutral hydrogen column density determined from the X-ray absorption and the true neutral hydrogen column density, we use the notation \nhx\ to refer to the former.

The results from our spectral analysis are provided in Table~\ref{tab:powsedfits} and Table~\ref{tab:bknpsedfits} for a power law and a broken power law fit, respectively. For five GRBs a broken power law provided little improvement to the fit with an F-test probability P $> 0.08$. For GRB~050802 and GRB~060418 a spectral break at 1--2~keV and $\sim 3$~keV, respectively, improved the goodness of fit of the model with an F-test probability ranging from 0.003 to $1.7\times 10^{-5}$ for GRB~050802 and from 0.02 to $9\times 10^{-4}$ for GRB~060418, depending on the model. However, the column density and the dust extinction at the host galaxy do not change significantly between the two continuum fits, remaining consistent to $2\sigma$.

For the purpose of this paper, where we are interested in the properties of the GRB local environment, the factors of importance are the dust extinction law that best fits the GRB afterglows and the relation between the column density and visual-extinction in the GRB local environment. We find that the former of these is not affected by the models that we use to fit the continuum, and there is no significant change in the ratio between the host galaxy column density and rest-frame visual extinction. This is illustrated in Fig.~\ref{fig:rat}, which plots the value of \nhx/\av\ for the best-fit parameters determined from a power law and a broken power law fit as solid and open circles, respectively. Beyond this section we, therefore, refer only to the spectral results from the power law fits. In Fig.~\ref{fig:seds} we show the SEDs and best-fit power law models for each GRB. In these figures data points at rest-frame wavelengths $\lambda < 1215$~\AA, which were not used in the fits (see section~\ref{sec:anlys}), are shown as open circles. 

The amount of intrinsic absorption and extinction determined in our sample varies by over a factor of five in \av\ and by more than an order of magnitude in \nhx. Typically the SMC model results in the smallest \av\ in the host galaxy and the MW model in the largest. This general trend is to be expected given the difference in FUV extinction observed in the three curves. The MW extinction law is the shallowest of the three extinction curves (see Fig.~\ref{fig:extcurves}) and, in particular, has the least amount of dust-absorption in the FUV for a given \av. A spectral model with a MW extinction law, therefore, requires a larger \av\ to fit the same data.

Although the X-ray column density and UV/optical extinction are independent components in the fit, the fitted value of \nhx\ depends on the spectral index, which in turn depends on the UV/optical extinction. The best-fit X-ray column density will, therefore, vary between extinction models. Typically the MW model requires a steeper spectral index to compensate for the reduced amount of FUV extinction, and this consequently results in larger absorption in the soft X-ray. Regardless of these differences the hydrogen equivalent column density determined from our spectral modelling is generally consistent at the $1\sigma$ level. In the case of GRB~050802 the column density is consistent at the $2\sigma$ level between dust models.

\begin{figure*}
\centering
\includegraphics[width=1.0\textwidth]{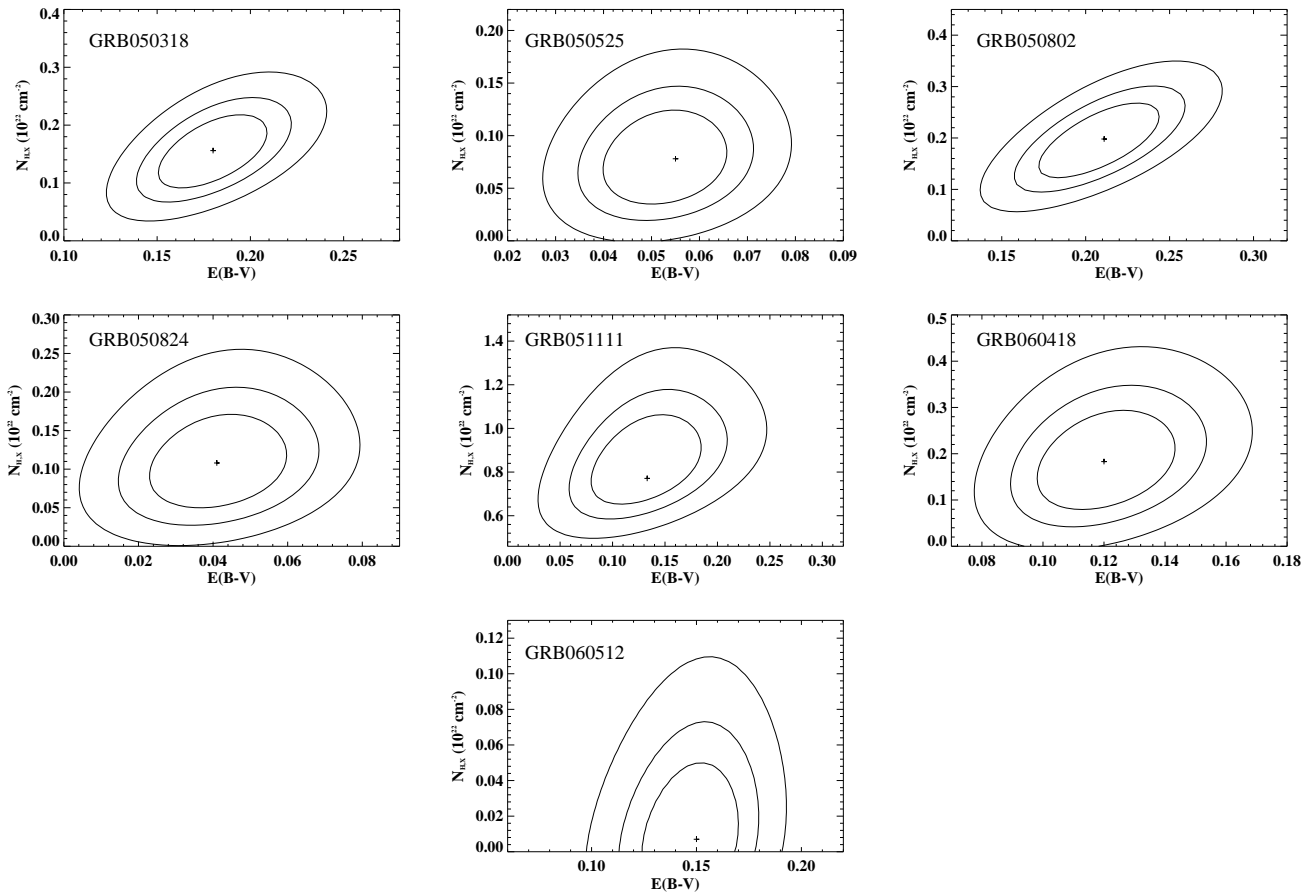}
\caption{Confidence contours for $E(B-V)$ vs. \nhx\ from the spectral model fits to the SEDs of the GRBs in the sample. For each GRB, the confidence contours shown are taken from the dust extinction model that provided the best fit. The contours are drawn at $\Delta\chi^2=2.3, 4.61, 9.21$, corresponding to 68\%, 90\% and 99\% confidence for two interesting parameters.}\label{fig:bfNHAVcontour}
\end{figure*}

\begin{figure}
\centering
\includegraphics[width=0.5\textwidth]{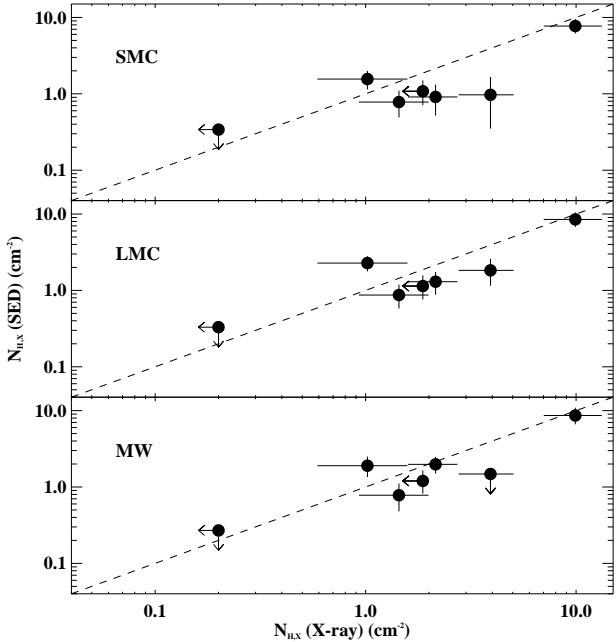}
\caption{Best-fit \nhx\ from SED spectral analysis vs. best-fit \nhx\ from spectral analysis to X-ray data alone. Top, middle and bottom panel correspond to the results from the SED spectral analysis using an SMC, LMC and MW model, respectively. In each panel the dashed line corresponds to \nhx (X-ray) = \nhx (SED). All data points lie on or very close to this line, illustrating the robustness of our spectral analysis.}\label{fig:NHvsNH}
\end{figure}

To further investigate the model dependence between the amount of X-ray absorption and dust extinction in the local environment of the GRB we show the confidence contours of $E(B-V)$ vs. \nhx\ for the best fit power law models to each GRB in Fig.~\ref{fig:bfNHAVcontour}. For the most part the contours are fairly circular, indicating that there is no significant correlation between \nhx\ and \av\ in our spectral modelling. We also tried fitting the X-ray data  alone to make sure that the UV/optical data were not in any way skewing the column densities. We find the best-fit \nhx\ value determined from our SED spectral analysis to be compatible with the best-fit \nhx\ value from spectral analysis on the X-ray data alone, as shown in Fig.~\ref{fig:NHvsNH}. This shows the robustness of our method.

\subsection{GRB Host Extinction Laws}
Six of the seven GRBs in our sample were best fit by the SMC or LMC model, with the MW model rejected with at least 97\% confidence for four of these (GRB~050318, 050525, 051111 and GRB~060512). For GRB~050824 and GRB~060418 the MW model is rejected with 84\% and 87\%, respectively. The relatively small amount of extinguishing dust in the circumburst environment of those two latter GRBs, indicated by the best fit \av, is likely to be the cause for the smaller distinction between the spectral models. Although the distinction in the goodness of fit between the SMC and LMC models is small, the SMC model provides the best fit to five of the GRBs in the sample, and only GRB~060418 is best fit by the LMC model; $\chi^2=68$ for 75 degrees of freedom (dof) compared to $\chi^2 = 79$ for 75 dof for the SMC model.

The afterglow of GRB~050525 was detected in all six lenticular UVOT filters \citep{bbb+06},  and GRB~050824 \citep{sc05} and GRB~060418 \citep{sf06} had an afterglow detection in all but the bluest UV filter ($UVW2$). GRB ~051111 did not have an afterglow detected in the two bluest filters \citep{psb+05}, and in the case of GRB~050318 \citep{srm+05} and GRB~060512 \citep{dc06} no afterglow was detected in any of the UV lenticular filters above the $3\sigma$ level.

At the redshifts of GRB~050318 [$z = 1.44$; \citet{bm05}], GRB~050824 [$z=0.83$; \citep{fjs+05}], GRB~051111 [$z = 1.549$; \citet{pro05}] and GRB~060418 [$z=1.49$; \citep{dfp+06}], the lack of an optical afterglow detection in the bluest filters could either be the result of Ly$\alpha$ blanketing or high levels of dust extinction blueward of $\sim 3800$~\AA. GRB~050318 and GRB~051111 have \av\ values that lie at the higher end of the distribution observed in the sample, with best fit parameters \av\ $=0.53\pm 0.06$~mag and \av\ $=0.39^{+0.11}_{-0.10}$~mag, respectively, if the SMC model is used. GRB~050824 and GRB~060418, on the other hand, have at least half this amount in rest-frame extinction if the SMC model is used (\av\ = $0.12\pm 0.04$ and \av\ = $0.17\pm 0.02$~mag, respectively). The correlation between \av\ and the number of filters in which the afterglow is detected could indicate that it is dust present in the local environment of these GRBs that contributes to the observed dimness of their UV afterglow.

GRB~060512 was a low redshift burst [z=0.4428; \citet{bfk+06}], eliminating neutral hydrogen absorption as the cause for the lack of an UV afterglow detection. The best fit is provided by the SMC model ($\chi ^2 =33$ for 20 dof) yielding a host galaxy extinction of \av\ $=0.44^{+0.04}_{-0.05}$~mag, which is comparable to that observed in GRB~050318 and GRB~051111. This, therefore, provides further support to the hypothesis whereby dust in the local environment of the GRB blocks a large fraction of the UV flux emitted.

In contrast to the other bursts discussed in this section, GRB~050525 had very small amounts of local absorption and extinction. However, good quality data resulting from the proximity [z=0.606; \citet{bbb+06}] and brightness of this burst in the UV and optical bands constrained well the spectral fits, and provided a distinction between them. The afterglow SED was best fit by the SMC model ($\chi ^2 = 29$ for 33 dof).

GRB~050802 is the only burst in our sample where the goodness of the spectral fit is improved with the MW model, ($\chi ^2 = 88$ for 81 dof). The SED of GRB~050802 flattens out at longer wavelengths (Fig.~\ref{fig:seds}) and this is well fit by a model with a dust extinction curve that contains the 2175~\AA\ absorption feature (\ie~LMC and MW models). The $\chi ^2$ of the LMC model fit is still acceptable, with $\chi ^2 = 98$ for 81 dof, although it is rejected at the 90\% confidence level in contrast to the MW model, which is only rejected at the 71\% confidence level.

\section{Discussion}\label{sec:disc}
\subsection{The 2175~\AA\ Absorption Feature}
The 2175~\AA\ absorption feature falls in the UVOT wavelength range for the GRB sample used in this paper, and its prominence in the GRB SEDs provides information on the graphite content of small grains in the surrounding circumburst material. The origin of the 2175~\AA\ feature is most likely to be carbonaceous material such as small spherical particles of graphite ($a\leq 30$~nm), which have a strong feature at this wavelength and of a similar width \citep{dl84}. Its strength in the MW extinction curve would require $\sim 15~\%$ of the solar abundance in carbon to be present in small particles of this size \citep{dra03}, whereas its absence in the SMC extinction curve can be explained by a difference in the relative abundances of graphite and silicate grains \citep{pei92}. 

The evidence for the 2175~\AA\ absorption dip in the SED of GRB~050802 suggests a larger abundance of small carbonaceous grains in the surrounding environment of this burst than is the case for the other bursts. A few other GRBs have also shown evidence for such a feature [\eg\ GRB~970508; \citep{sfa+04,kkz06}, GRB~991216; \citep{vsf+06}], although the usual absence of this feature in the spectra of GRBs indicate it to be rare in GRB host galaxies.

\subsection{X-ray Absorption vs. UV/Optical Extinction}\label{subsec:NHAV}
\begin{figure*}
\centering
\includegraphics[width=1.0\textwidth]{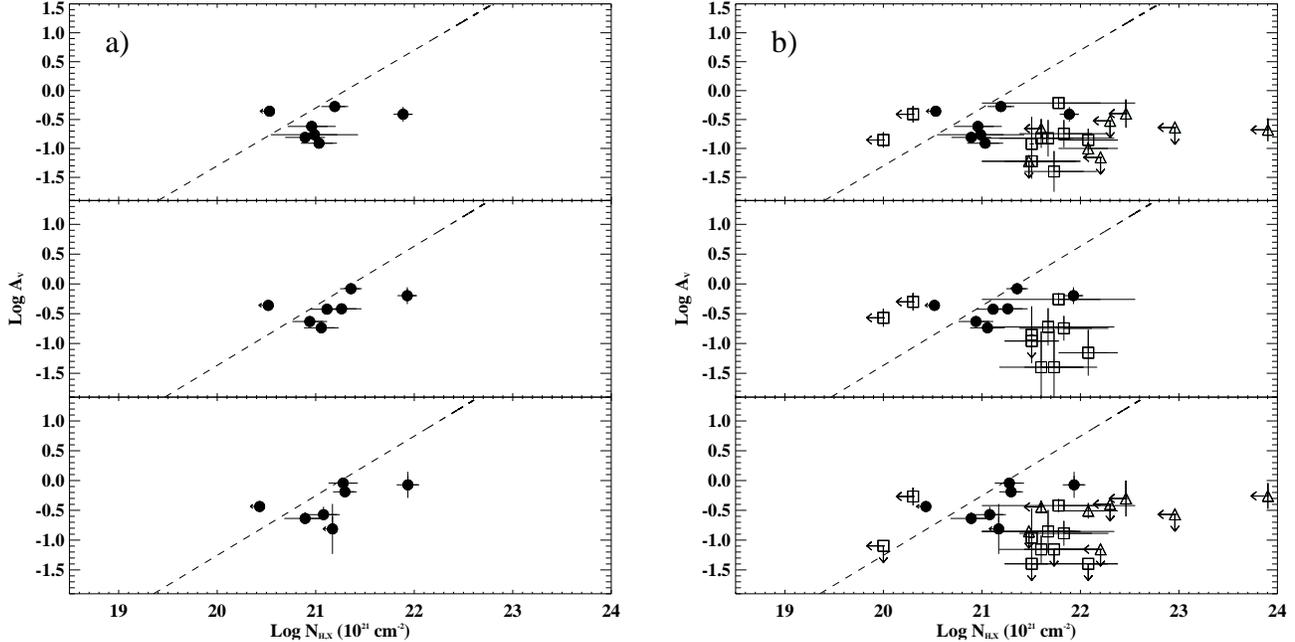}
\caption{Host galaxy \av\ vs. \nhx. In both plots solid circles represent the host galaxy \av\ and \nhx\ determined from our spectral analysis using the SMC model (top panel), the LMC model (middle panel) and the MW model (bottom panel). The dashed curves are the \nhx/\av\ ratios for each corresponding environment. This was determined using the \nh/\av\ ratios reported in the literature, where $N_H$ is converted to the X-ray equivalent \nhx\ value assuming a metallicity 1/8 and 1/3 solar for the SMC and LMC, respectively (see text for details). In the right panel the open triangles and open squares are GRBs taken from the SFA and KKZ sample, respectively.}\label{fig:AVNH}
\end{figure*}

The amount of dust extinction observed in the afterglow of GRBs is a measure of the column density of dust grains responsible for the absorption of UV and optical photons, and the ratio between the X-ray column density and extinction gives an estimate of the gas-to-dust ratio in the surrounding environment of the GRB. In Fig.~\ref{fig:AVNH}a we show the region of \av\ and \nhx\ parameter space occupied by our sample of bright GRBs for the best fit \av\ and \nhx\ values determined from the spectral analysis when fitting the SMC (top panel), the LMC (middle panel), and the MW models (bottom panel) to the data.

The dashed lines in the figure are plotted as a point of reference and correspond to the empirical relation observed between \av\ and \nhx\ in the SMC (top), LMC (middle) and MW (bottom). These are determined from the \nh/\av\ values reported in the literature for each of these environments, which are then converted to an \nhx/\av\ ratio relating to the column density that would be measured from X-ray observations of the galaxy if solar abundances were assumed. The parameterisation of these lines differ in each panel, and correspond to
\begin{eqnarray}
\frac{N_{H.X}}{A_{V}}\rm{(SMC)} =& \frac{1}{8}\frac{N_{H}}{A_{V}}\rm{(SMC)} = \frac{1}{8}(1.6\times 10^{22})~\rm{cm}^{-2}\label{eqn:NHAVrat1}\\
\frac{N_{H,X}}{A_{V}}\rm{(LMC)}  =& \frac{1}{3}\frac{N_{H}}{A_{V}}\rm{(LMC)} = \frac{1}{3}(0.7\times 10^{22})~\rm{cm}^{-2}\label{eqn:NHAVrat2}\\
\frac{N_{H,X}}{A_{V}}\rm{(MW)}  =& \frac{N_{H}}{A_{V}}\rm{(MW)} = 0.18\times 10^{22}~\rm{cm}^{-2}\label{eqn:NHAVrat3}
\end{eqnarray}
where the factors of $\frac{1}{8}$ and $\frac{1}{3}$ account for the lower metallicities observed in the SMC and LMC \citep{pei92}. The \nh/\av\ relations are taken from \citet{wed00} and \citet{ps95} for the SMC and MW, respectively, and the average of the ratios found by \citet{koo82} and \citet{fit85} are used for the LMC.

The data points in Fig.~\ref{fig:AVNH} primarily lie to the right of the lines of constant \nhx/\av. However, they are confined to an area of the \nhx--\av\ parameter space much closer to the dust-to-gas ratio observed in the MW and Magellanic Clouds than the region of space occupied by previous data. This is illustrated in Fig.~\ref{fig:AVNH}b, where we include the host galaxy \nhx\ and \av\ for two pre-\swift\ GRB samples taken from \citet{sfa+04} (SFA sample) and \citet{kkz06} (KKZ sample), which are shown as open triangles and open squares, respectively. Despite apparent differences between the \swift\ sample and previous samples, the distribution in \av\ and \nhx\ between samples remain consistent within errors. \citet{sfa+04} did not use the LMC extinction law to model the afterglow spectrum, and consequently there are no triangles shown in the centre panel of Fig.~\ref{fig:AVNH}b.

The X-ray data in the SFA sample are from {\it BeppoSAX} and were taken hours to days after the prompt outburst, with the earliest observation beginning at $\sim \rm{T}+4$~hours. In contrast to this, the GRBs in our sample typically have X-ray data starting hundreds of seconds after the BAT trigger, and the longest delay between the GRB prompt emission and the first X-ray observations is 1.7~hours \citep[GRB~050824; ][]{cmc+05}. The larger signal-to-noise provided at early times when the afterglow is significantly brighter improves the accuracy of our spectral analysis and reduces the errors associated with the best fit parameters. \citet{sfa+04} point out that the quality of the X-ray data for the majority of the bursts in their sample does not allow for significant detections of host galaxy absorption in addition to Galactic, and in only the cases of GRB~990123 and GRB~010222 were the presence of excess absorption robustly detected. This is indicated in Fig.~\ref{fig:AVNH}b, where the \nhx\ value for all but two of the SFA sample are upper limits. 

\citet{kkz06} focused primarily on optical and NIR data, where X-ray column densities used in their analysis were taken from the literature. In their sample the earliest X-ray observation was still only $\sim\rm{T} + 4$~hours and the average delay was nearly 50~hours from the time of initial outburst. However, for most of their sample the column density was determined from higher quality X-ray data taken with {\it XMM-Newton} or {\it Chandra}. For this sample the host galaxy \nhx\ ranged from an undetectable amount \citep[GRB~021004;][]{mfh+02} to $(12^{+7}_{-6})\times 10^{21}$~\invsqrcm\ \citep[GRB~010222;][]{sfa+04}, and the mean is $(4.6^{+1.7}_{-1.0})\times 10^{21}$~\invsqrcm, which is consistent within $2\sigma$ of our sample, which is $(2.2\pm 0.3)\times 10^{21}$~\invsqrcm. Their optical and NIR analysis of 19 GRBs provided a distribution in \av\ that ranged from a negligible amount up to \av = $0.80\pm 0.29$~mag, which is in good agreement with our results, for which \av\ ranges from \av\ = $0.12\pm 0.04$ to \av\ = $0.65^{+0.08}_{-0.07}$~mag. The mean extinction is $0.21\pm 0.04$~mags and $0.38\pm 0.02$~mags for the KKZ and our sample, respectively.

The high quality of the data in our sample and the simultaneous spectral fitting of the X-ray and UV/optical data is providing greater constraints on the spectral modelling, and consequently reducing the errors on the best-fit parameters. The results from this are that the gas-to-dust ratios in the local environment of GRBs are in better agreement with those observed in the Milky Way and Magellanic Clouds than previous data suggest. The relatively large gas-to-dust ratios in GRB local environments indicated by previous data were interpreted as evidence of dust destruction by the GRB, which would cause the value of \av\ to decrease. However, the results from the \swift\ data analysis show little evidence of this.

Moreover, when interpreting the \nhx\ to \av\ ratio in the GRB local environment compared with that of other environments, it is necessary to consider more than just the effect of the GRB emission on \av. Photoionisation of the gas in the surrounding environment by the GRB X-ray radiation causes \nhx\ to decrease with time, and the extent to which the GRB reduces the \av\ and \nhx\ will depend on the properties of the GRB and its local environment, such as the prompt and afterglow spectral and temporal indices, and the density and density profile of the absorbing and extinguishing material.

The lack of evidence for any colour evolution in our UV/optical data (see section~\ref{subsec:opt}), provides an upper limit on the time by which the GRB no longer destroyed significant amounts of dust. \ie\ the time at which colour information is available with UVOT, which is typically within $\sim 10^3$~s. A limit on the time after which the GRB no longer photoionises the surrounding environment can also be determined by investigating the change in \nhx\ over time during the early stages of the X-ray afterglow. For this purpose, we applied spectral analysis on the early time X-ray data of the five GRBs in our sample for which there were data within the first few 100 seconds of the BAT trigger. Of these GRBs only GRB~060418 shows evidence for evolution in the column density at greater than the $3\sigma$ level. The column density measured from T+84~s to T+114~s was $(1.6\pm 0.10)\times 10^{22}$~\invsqrcm, whereas beyond T+400~s this was $(3.9^{+1.3}_{-1.2})\times 10^{21}$~\invsqrcm. GRB~060418 had a very large X-ray flare that peaked at T+135~s \citep{fbm+06} and increased the X-ray flux by about an order of magnitude in $\sim 15$~s, which could have caused photoionisation of the circumburst environment out to greater radii from the source. However, it is possible to misinterpret intrinsic spectral evolution as a change in the column density \citep{bk06}. The spectral index indeed changed from $\Gamma=2.75\pm 0.05$ to $\Gamma=2.15\pm 0.08$ between the two spectral epochs analysed, and the apparently larger \nhx\ could be due to intrinsic curvature of the flare spectrum. The column density measured in GRB~060418 no longer evolves beyond T+300~s, by which time the flare is over and, therefore, we determine a typical upper limit of a few hundred seconds on the time interval over which the GRB photoionises its surrounding environment.

\citet{pl02} simulated the effect of the GRB X-ray and UV emission on \nhx\ and \av\ over time, where they assumed the same initial hydrogen column density $N_H=10^{22}$~\invsqrcm\ and optical extinction $A_{V}=4.5$~mag in all cases, but varied the compactness of the absorbing medium and, therefore, also the number density, $n_H$. Their simulations indicate that the intense radiation emitted by a GRB is capable of photoionising and destroying all gas and dust out to a radius of $\sim 3$~pc within a few tens of seconds. Further evolution in the gas and dust column resulting from this is, therefore, not expected more than a few hundreds of seconds after the peak of the emission, consistent with our observations.

This, therefore, places a lower limit of a few parsecs on the scale of the absorbing and extinguishing systems detected at the redshifts of GRBs for six of the seven in our sample. If the dust and gas were any closer it would have been fully destroyed and photoionised. Furthermore, \citet{pl02}'s work shows that if the dust and gas extends out to a few tens of parsecs the \av\ and \nhx\ measured local to the GRB do not change significantly. This would suggest that the host galaxy dust and gas systems probed by our sample lie a few tens of parsecs from the source, which is consistent with findings by Prochaska, Chen \& Bloom (2006), who placed a lower limit of $\gtrsim 50$~pc on the locations of host galaxy absorption systems from GRBs. This suggests that the dust-to-gas ratios measured are effectively unaltered by the GRB and a fair representation of their local environment. It is, therefore, not so surprising that there is no significant deviation in the GRB local environment \nhx/\av\ ratio when compared to the Milky Way or Magellanic Clouds.

To verify that the effect of the GRB on the dust and gas in its local environment is of little significance, we investigate further alternative methods of indirectly detecting the process of dust destruction.

\begin{figure}
\centering
\includegraphics[width=0.5\textwidth]{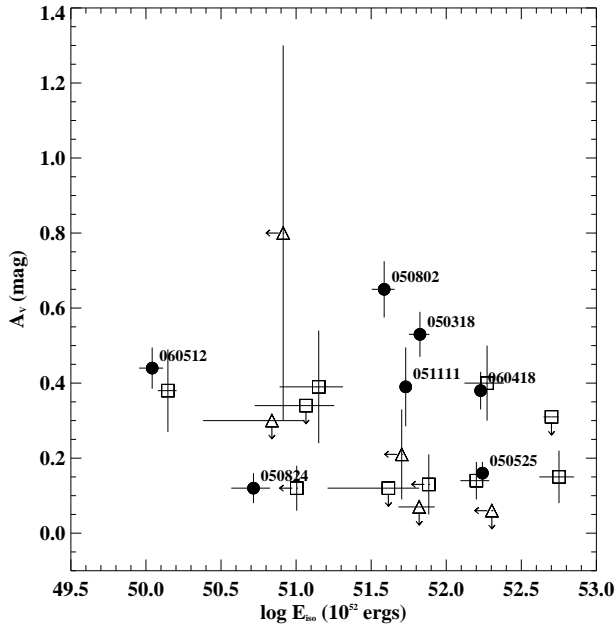}
\caption{Host galaxy visual extinction against isotropic equivalent energy in $\gamma$-rays for a sample of 16 bursts. Solid data points correspond to the GRBs studied in this paper. Open data points are GRBs from a pre-\swift\ sample of GRBs where the \Eiso\ values are taken from \citet{bfs01} and the \av\ values are from \citet{kkz06} (square) and \citet{sfa+04} (triangle).}\label{fig:EvsAv}
\end{figure}

\subsection{Alternative Indicators Of Dust Destruction}\label{subsec:dd}
Dust destruction, either by sublimation or shattering is dependent on both the energy of the photons absorbed and on the radiation flux, where a less-luminous burst will be less effective at destroying the smaller, UV-absorbing dust grains. If the radiation emitted by a GRB destroys significant levels of dust in its surrounding environment a correlation should exist between the energy released by the burst and the amount of dust destroyed, which consequently affects the amount of extinction observed in the GRB afterglow spectrum.

To test this correlation we compute the isotropic equivalent energy emitted in the 15 -- 150~keV $\gamma$-ray energy band (\Eiso $(\gamma)$) for our sample of GRBs. A k-correction is applied using the method described by Bloom, Frail \& Sari (2001) to convert the prompt energy observed to the same comoving rest frame bandpass (15 -- 150~keV). Fig.~\ref{fig:EvsAv} shows the host galaxy \av\ against the k-corrected \Eiso $(\gamma)$ for the seven GRBs in our sample (solid circles), as well as a further fifteen pre-\swift\ GRBs (open squares or triangles). All the GRBs included in Fig.~\ref{fig:EvsAv} have spectroscopically measured redshifts and estimates of the host galaxy \av, where the host galaxy \av\ for the pre-\swift\ sample are taken from either \citet{kkz06} (open squares) or \citet{sfa+04} (open triangles). A similar k-correction is applied to the non-\swift\ GRB sample to determine \Eiso $(\gamma)$ in the same rest-frame energy range as the \swift\ sample.

Twelve of the pre-\swift\ GRBs had no spectral information and, therefore, the spectral shape had to be assumed. \citet{bmf+93} found the GRB prompt emission to be well modelled by a smoothed, broken, power law with a soft and hard spectral index, $\alpha$ and $\beta$, respectively, and spectral break $E_0$. Those GRBs with no spectral information are assumed to have a Band spectrum, and their \Eiso $(\gamma)$ is taken from the average of 54 values, each computed with a different set of ($\alpha$,$\beta$,$E_0$). The set of 54 ($\alpha$,$\beta$,$E_0$) combinations are taken from fits to 54 BATSE GRBs done by \citet{bmf+93}.

Based on the Spearman's rank test the correlation in Fig.~\ref{fig:EvsAv} is not significant, with the log of \Eiso\ and \av\ showing a correlation coefficient of only $r_{s}=-0.21$ with a null-hypothesis probability of P=0.60. This is consistent with the findings of \citet{ngg+06}, whose analysis showed a sample of 23 pre-\swift\ GRBs to show no correlation between the isotropic $\gamma$-ray emitted energies and the optical luminosities. This, therefore, suggests that any dust-destruction caused by the GRB intense radiation is not significant enough to affect greatly their observed extinction properties, as already indicated in Fig.~\ref{fig:AVNH} and discussed in section \ref{subsec:NHAV}.

\subsection{Extension to Dark Bursts}\label{sec:darkbrsts}
By imposing the criteria that the selected GRBs have a UV/optical afterglow detected by UVOT, a bias is introduced that favours afterglows with low source-frame extinction. The determined range in host-galaxy extinction may, therefore, represent the low end of the extinction distribution. Given the association of GRBs with star formation, it is likely that the lack of an optical counterpart in some bursts is due to large amounts of dust in the host galaxy, which blocks out the UV/optical afterglow. In this case the distribution of host galaxy \av\ may extend much further than that which is observed in our sample. GRBs with no associated optical counterpart are typically referred to as `dark', although the various possible causes for the lack of an optical counterpart make the term ambiguous \citep{rwk+05,rm06}. In this paper we shall use the term dark to refer to those GRBs that have no afterglow detected at the $3\sigma$ level above background in any of the UVOT filters within one hour of the prompt emission.

\begin{figure}
\centering
\includegraphics[width=0.5\textwidth]{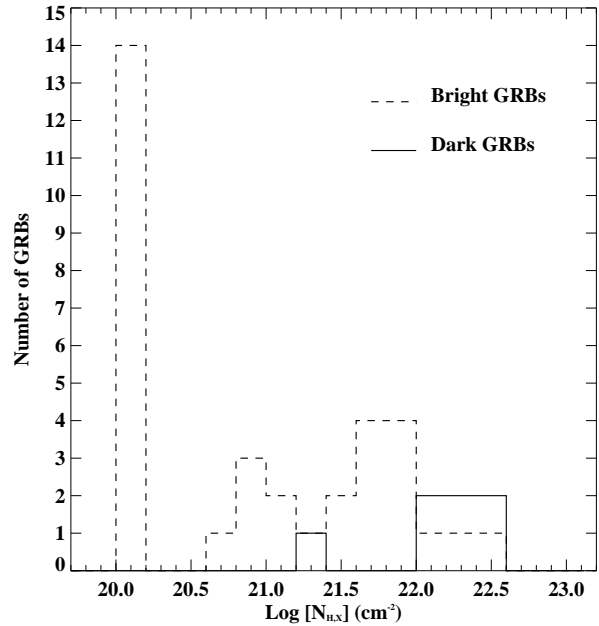}
\caption{Distribution of the X-ray column density at the redshift of the GRB for a sample of GRBs with (dashed), and without (solid) a UVOT detected afterglow, where all GRBs have a spectroscopically measured redshift. The smallest \nhx\ bin represents those GRBs with no measurable X-ray column density above the Galactic absorption.}\label{fig:NHdist}
\end{figure}

There are seven \swift\ GRBs that satisfy both our definition of dark and which have spectroscopically measured redshifts, where $z < 5$. At higher redshifts than this a GRB would always appear dark to UVOT due to the redshifting of the Lyman break below the UVOT energy band. To estimate the amount of dust in the local environment of this sample of GRBs, we initially assume the relation between \nhx\ and \av\ to be linear, and thus use the X-ray column density, \nhx, of the GRB host galaxy to trace the amount of dust-extinction affecting the GRB UV/optical afterglow. We compare this with the dust extinction estimated for a sample of 34 optically bright \swift\ detected GRBs. This sample is made up of those GRBs with an UV/optical afterglow detected by UVOT and a corresponding spectroscopic redshift, up until GRB~060912; this includes our original sample of seven UVOT bright GRBs listed in table~\ref{tab:powsedfits}. All data were taken from epochs where no spectral evolution is observed, and it was reduced in the same way as described in section~\ref{subsec:xray}. The intrinsic column density was then determined from a single power law fit with two absorbers; one at $z=0$ with the column density fixed at the Galactic value, and a second one at the redshift of the burst with the column density and X-ray spectral slope, $\beta_X$, left as free parameters. The resulting distribution in \nhx\ within the two populations of bursts is shown in Fig.~\ref{fig:NHdist}, where the solid histogram corresponds to the dark bursts, and the UVOT bright bursts are represented by the dashed histogram. Fourteen bright GRBs had negligible absorption at the host galaxy, represented in the smallest bin of \nhx. The mean logarithm of the X-ray column density, over and above the Galactic column is $< 20.54$~\invsqrcm\ and $22.2\pm 0.1$~\invsqrcm\ for the bright and dark population of GRBs, respectively, suggesting that dark GRBs reside in denser environments.

Assuming GRB host galaxies to have an \nhx\ and \av\ relation similar to that observed in the SMC (Eqn.~\ref{eqn:NHAVrat1}), we use the measured \nhx\ to determine \av. Doing this we estimate a mean visual extinction of $\langle A_{V} (\rm{dark})\rangle = 10.1$~mags for the dark population of bursts shown in Fig.~\ref{fig:NHdist}. The most absorbed GRB in the sample of dark GRBs was GRB~060510B, which had a rest-frame column density \nhx = $(3.54^{+0.24}_{-0.23})\times 10^{22}$~\invsqrcm\ and, therefore, an estimated rest-frame visual extinction of \av$\approx 17.7$~mags. However, GRB~060510B had an R-band detection of R~$\sim 21$ at around 12~minutes after the prompt emission, making the inferred extinction rather unrealistic, which suggests that the \nhx/\av\ ratio of $2\times 10^{21}$~\invsqrcm\ is likely too small for the most absorbed of GRBs.

Instead, we use the host galaxy gas-to-dust ratios determined in our spectral analysis to acquire a mean value of \nhx/\av, where we use the values determined from the SMC model. This corresponds to:
\begin{eqnarray}
\langle\frac{N_{H,X}}{A_{V}}\rangle = 6.7\times 10^{21}~\rm{cm}^{-2}\label{eq:rat4}
\end{eqnarray}
Using this relation we estimate a range in rest-frame \av\ of $0.4 < A_V < 5.3$, with a mean of $\langle A_{V}\rangle =  3.0$~mags. The overlap in rest-frame visual extinction in the UVOT dark and bright population of GRBs suggests that several factors may contribute to the lack of an optical afterglow. However, the mean value of $\langle A_{V}\rangle =  3.0$~mags is almost eight times the mean extinction observed in the bright GRBs listed in Table~\ref{tab:powsedfits}, indicating that the local environments of UVOT dark GRBs are much dustier than those with UVOT detected optical counterparts. This could explain, or at least contribute, to the lack of an afterglow blueward of 5500~\AA. 

\section{Conclusions}\label{sec:conc}
In this paper we use the SEDs of seven optically bright GRBs, covering the optical, UV and X-ray energy bands, to determine the amount of dust extinction and soft X-ray absorption present in the local environments of the GRB. We use the Milky Way (MW) and the Large and Small Magellanic Cloud (LMC and SMC) extinction laws to model the host galaxy dust extinction dependence on wavelength. These show a decreasing prominence in the strength of the 2175~\AA\ feature, and increasing levels of far-ultraviolet extinction, respectively.

All GRBs but one were best-fit by the SMC or LMC extinction law, with the SMC extinction law typically giving a better fit. Only in the case of GRB~050802 did a model with a Galactic extinction law at the host galaxy provide a better fit, although SMC and LMC extinction laws were rejected with only 97\% and 90\% confidence, respectively. This, therefore, suggests that small graphite grains responsible for the 2175\AA\ feature are not predominant in GRB host galaxies.

For six of the seven GRBs in our sample an absorption and extinction system was detected at the redshift of the GRB, which must be located at least several parsecs from the source to have survived the intense radiation emitted by the GRB. However, the gas-to-dust ratios measured in the host galaxies of our sample of bursts are lower than previously suggested from analysis of pre-\swift\ GRBs, and consistent with those observed in the Milky Way and Magellanic Clouds. There is, therefore no evidence of dust destruction by the GRB in its circumburst environment provided by the spectral analysis alone. However, this does not rule out the destruction and photoionisation of dust and gas within a few parsecs of the GRB, which is expected to take place within the first few tens of seconds of the GRB prompt emission.

The host galaxy visual extinction observed in our sample ranges from \av\ $=0.12\pm 0.04$ to $0.65^{+0.08}_{-0.07}$~mags. However, the requirement that the GRBs analysed have a detected optical counterpart introduces selection effects that favour GRBs with a low \av. Using the amount of host galaxy \nhx\ as an indicator of the amount of visual extinction undergone in the GRB local environment, we estimate that those GRBs with no optical counterpart have, on average, values of \av\ almost a factor of eight larger than those observed in the sample of bright GRBs studied in this paper. This would suggest that the host galaxies of dark GRBs are intrinsically dustier than those of GRBs with UVOT detected optical counterparts, which could account for the lack of an optical afterglow.

\section*{ACKNOWLEDGEMENTS}
We gratefully acknowledge the contributions of all members of the {\it Swift} team. PS acknowledges the support of a PPARC Studentship. PS would also like to thank David Morris and Claudio Pagani for all their patience and help regarding the reduction of XRT data.

\end{document}